\documentclass[aps,prx,twocolumn,superscriptaddress,longbibliography]{revtex4-1}

\usepackage{amsmath}
\usepackage{amssymb}
\usepackage{graphicx}
\usepackage[dvipsnames]{xcolor}
\usepackage[colorlinks,linkcolor=blue,citecolor=blue,urlcolor=blue]{hyperref}
\usepackage{stix}

\begin{document}

\renewcommand{\vec}[1]{\mathbf{#1}}
\newcommand{\iu}{\mathrm{i}}
\newcommand{\hc}{\hat{c}}
\newcommand{\hcd}{\hat{c}^\dagger}
\newcommand{\en}{\varepsilon}
\newcommand{\gvec}[1]{\boldsymbol{#1}}
\renewcommand{\pl}{\parallel}

\newcommand{\new}[1]{\textcolor{WildStrawberry}{#1}}

\author{Michael Sch\"uler}
\email{michael.schueler@psi.ch}
\affiliation{Stanford Institute for Materials and Energy Sciences (SIMES),
  SLAC National Accelerator Laboratory, Menlo Park, CA 94025, USA}
\affiliation{Condensed Matter Theory Group, Paul Scherrer Institute, CH-5232 Villigen PSI, Switzerland}
\author{Tommaso Pincelli}
\affiliation{Fritz Haber Institute of the Max Planck Society, Faradayweg 4-6, 14195 Berlin, Germany}
\author{Shuo Dong}
\affiliation{Fritz Haber Institute of the Max Planck Society, Faradayweg 4-6, 14195 Berlin, Germany}
\author{Thomas P. Devereaux}
\affiliation{Stanford Institute for Materials and Energy Sciences (SIMES),
  SLAC National Accelerator Laboratory, Menlo Park, CA 94025, USA}
\affiliation{Department of Materials Science and Engineering, Stanford University, Stanford, California 94305, USA}
\author{Martin Wolf}
\affiliation{Fritz Haber Institute of the Max Planck Society, Faradayweg 4-6, 14195 Berlin, Germany}
\author{Laurenz Rettig}
\affiliation{Fritz Haber Institute of the Max Planck Society, Faradayweg 4-6, 14195 Berlin, Germany}
\author{Ralph Ernstorfer}
\email{ernstorfer@fhi-berlin.mpg.de}
\affiliation{Fritz Haber Institute of the Max Planck Society, Faradayweg 4-6, 14195 Berlin, Germany}
\author{Samuel Beaulieu}
\email{samuel.beaulieu@u-bordeaux.fr}
\affiliation{Fritz Haber Institute of the Max Planck Society, Faradayweg 4-6, 14195 Berlin, Germany}
\affiliation{Université de Bordeaux - CNRS - CEA, CELIA, UMR5107, F33405, Talence, France}

\title{Polarization-Modulated Angle-Resolved Photoemission Spectroscopy: \\ 
Towards Circular Dichroism without Circular Photons and Bloch Wavefunction Reconstruction}


\begin{abstract} Angle-resolved photoemission spectroscopy (ARPES) is the most powerful technique to investigate the electronic band structure of crystalline solids. To completely characterize the electronic structure of topological materials, one needs to go beyond band structure mapping and access information about the momentum-resolved Bloch wavefunction, namely orbitals, Berry curvature, and topological invariants. However, because phase information is lost in the process of measuring photoemission intensities, retrieving the complex-valued Bloch wavefunction from photoemission data has yet remained elusive. We introduce a novel measurement methodology and associated observable in extreme ultraviolet angle-resolved photoemission spectroscopy, based on continuous modulation of the ionizing radiation polarization axis. Tracking the energy- and momentum-resolved amplitude and phase of the photoemission intensity modulation upon polarization axis rotation allows us to retrieve the circular dichroism in photoelectron angular distributions (CDAD) without using circular photons, providing direct insights into the phase of photoemission matrix elements. In the case of two relevant bands, it is possible to reconstruct the orbital pseudospin (and thus the Bloch wavefunction) with moderate theory input, as demonstrated for the prototypical layered semiconducting transition metal dichalcogenide 2H-WSe$_2$. This novel measurement methodology in ARPES, which is articulated around the manipulation of the photoionization transition dipole matrix element, in combination with a simple tight-binding theory, is general and adds a new dimension to obtaining insights into the orbital pseudospin, Berry curvature, and Bloch wavefunctions of many relevant crystalline solids.

\end{abstract}

\maketitle


\section{Introduction}

Wavefunctions are mathematical descriptions of the quantum state of a system and are ubiquitous in quantum mechanics. They are complex-valued probability amplitudes, and the probabilities for the results of any measurements made on a quantum system can be derived from them. Because of their complex-valued nature, and since most experimental techniques are only sensitive to the square-modulus of the wavefunction -- leading to a loss of the phase information -- reconstructing wavefunctions from experimental observables is a challenging task. 

The use of interferometric measurement techniques, which use the interference pattern generated by superimposed waves to extract their relative phases, has been used to experimentally reconstruct the electronic wavefunction of atoms and molecules. For example, the interferometric nature of the photoelectric effect, as well as its time-reversed analog \textit{photorecombination}, have been used to reconstruct the orbitals of atoms~\cite{Shafir_09,Villeneuve_17}, aligned gas-phase molecules~\cite{Itatani_04,Haessler_10}, as well as molecular adsorbates~\cite{Puschnig_09,Wiessner_14}. Real-space excitonic wavefunction has also recently been reconstructed using the Fourier transform of the momentum-space photoemission intensity~\cite{man20,dong_direct_2021}, assuming a flat phase. 

Knowledge about the electronic band structure, \textit{i.e.} the momentum-dependent energy eigenvalues, and the associated Bloch wavefunction are essential to understand the transport, optical and magnetic properties of crystalline solids. With the discovery of topological materials~\cite{Kane_05}, it became clear that accessing knowledge beyond band structure is of fundamental importance to understand the unique properties of this important class of quantum materials. The topologically non-trivial nature of materials emerges from the winding of the phase of their Bloch wavefunctions in momentum-space, associated with Berry curvature~\cite{Berry_84} and topological invariants, \textit{e.g.} Chern numbers~\cite{hasan_colloquium:_2010-1,qi_topological_2011}. Reconstructing the band structure \textit{and} the associated Bloch wavefunction is thus of capital importance to fully characterize the electronic structure of (topological) materials. 

While the electronic band structures of crystalline materials can be mapped using angle-resolved photoemission spectroscopy (ARPES)~\cite{Damascelli_04,gedik_photoemission_2017,lv_angle-resolved_2019,sobota_angle-resolved_2021}, reconstructing the associated Bloch wavefunction is still a great challenge. Whereas complex-valued information about the Bloch wavefunction of electrons inside solids is encoded in the photoionization transition dipole matrix element underlying the photoelectric effect, leading to subtle anisotropic modulation of the signal in momentum-energy space, a general route to reconstruct the Bloch wavefunction from photoemission data have not been established yet. 

Circular dichroism in the photoelectron angular distribution (CDAD) is a powerful quantity that can be used to probe \textit{e.g.} electronic chirality in graphene~\cite{Liu11}, helical spin textures in topological insulators~\cite{Wang11,lin_orbital-dependent_2018,jozwiak_spin-polarized_2016}, the orbital Rashba effect in metals~\cite{Park12}, high-symmetry planes~\cite{fedchenko_4d_2019}, and the Berry curvature in TMDCs~\cite{razzoli_selective_2017,cho_experimental_2018,cho_studying_2021, schuler_local_2020-1}. In contrast, linear dichroism in the photoelectron angular distribution (LDAD) is typically assumed to encode the non-relativistic symmetry of the wavefunction~\cite{Schonhense90,Cherepkov93,Sterzi18,Volckaert19,rostami_layer_2019,beaulieu2021unveiling}, but does not contain enough information to access the phase of the Bloch wavefunction. 

Here, we introduce a novel measurement scheme in extreme ultraviolet (XUV) angle-resolved photoemission spectroscopy, based on a continuous rotation of the polarization axis, allowing to reconstruct: i) CDAD without using circular photons, and ii) the phase of the photoemission matrix elements, which directly relates to the complex-valued Bloch wavefunction, here exemplified for 2H-WSe$_2$. This information can, in principle, even be used to reconstruct the orbital pseudospin texture. Indeed, tracking the energy- and momentum-resolved modulation of the photoemission intensity upon continuous rotation of the ionizing radiation polarization axis, complemented by theory input, enable us to go beyond band structure mapping and access properties of the Bloch wavefunction underlying the electronic band structure of crystalline solids.

For the first demonstration of our novel approach, we choose to study the layered transition metal dichalcogenide (TMDC) 2H-WSe$_2$. Despite its inversion-symmetric crystal structure, this material possesses locally broken inversion symmetry within each layer and strong spin-orbit coupling, leading to entangled layer, spin, orbital, and valley degrees of freedom \cite{Zhang14}. The topmost layer surface sensitivity of XUV-ARPES allows to directly probe this intricate hidden spin \cite{Riley14,razzoli_selective_2017} and orbital \cite{beaulieu_revealing_2020-1} texture. This peculiar spin-orbital-valley locking leads to optical selection rules allowing for the generation of spin- and valley-polarized excited carriers~\cite{Bertoni16}, to orbital Hall effect (OHE) \cite{Go18}, and the emergence of orbital Hall insulating phases \cite{Canonico20,cysne_disentangling_2021}. The valley-dependent orbital pseudospin texture is also at the origin of the emergence of local Berry curvature~\cite{Berry_84}, associated with the winding of the wavefunction phase in momentum-space~\cite{cao_unifying_2018}. Such material is thus well suited to test our novel polarization-modulated angle-resolved photoemission spectroscopy approach.

\section{Results}

\begin{figure}[t]
\centering\includegraphics[width=\columnwidth]{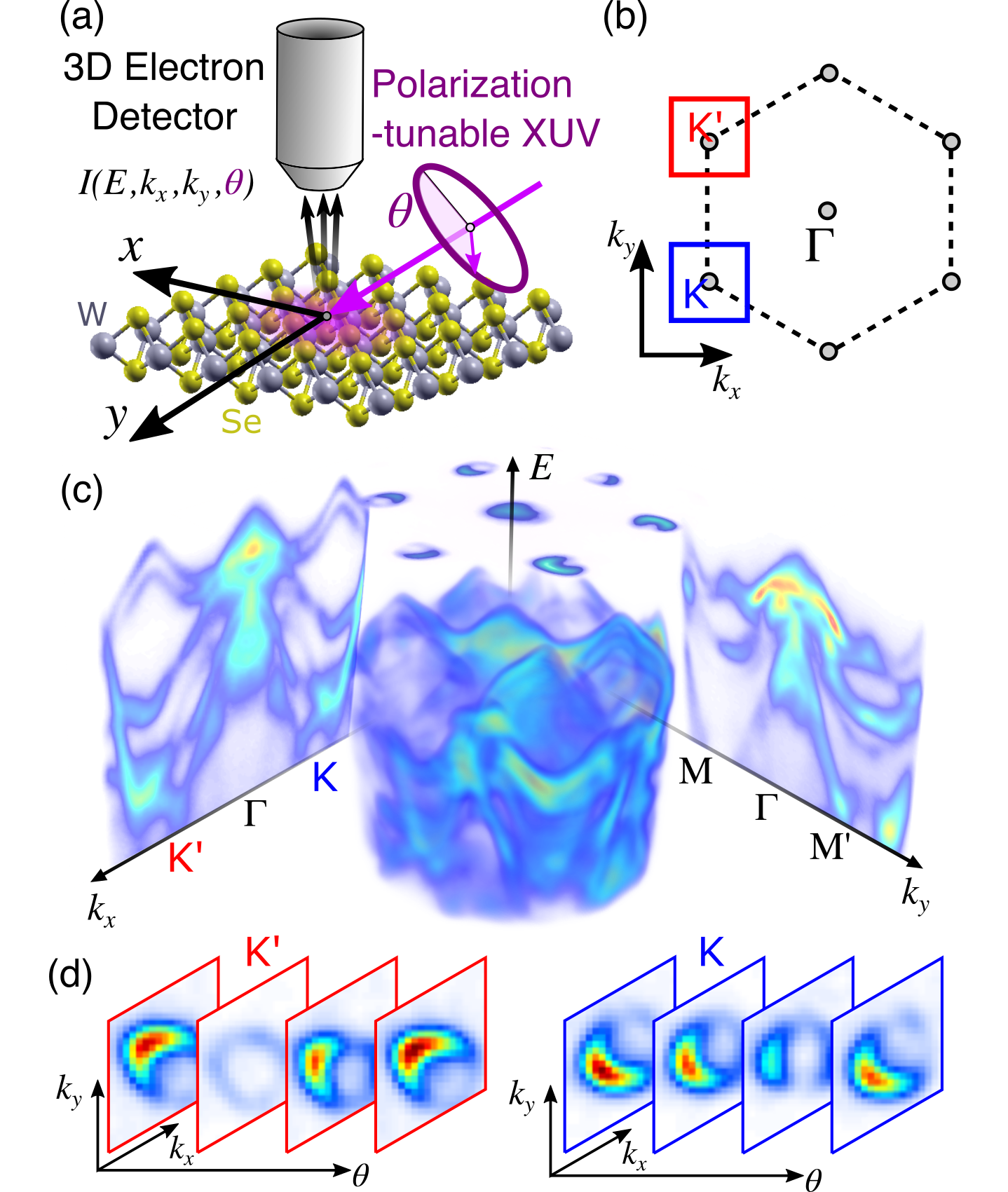}
\caption{\textbf{Experimental setup and measurement protocol:} (a) Experimental scheme of polarization-modulated angle-resolved photoemission spectroscopy. A polarization-axis-tunable linearly polarized femtosecond XUV pulse (21.7 eV) is focused onto a bulk 2H-WSe$_2$ crystal at an angle of incidence of 65$^{\circ}$ with respect to the surface normal, ejecting photoelectrons which are detected by a time-of-flight momentum microscope, allowing to measure the energy- and momentum-resolved photoemission intensity as a function of the polarization axis angle $\theta$ - $I(E,k_x,k_y,\theta)$. (b) Sketch of the first Brillouin zone of 2H-WSe$_2$. (c) Example of three-dimensional raw data -  band structure mapping ($I(E,k_x,k_y)$) using p-polarized XUV radiation, associated cut through high-symmetry directions (K$^\prime$-$\Gamma$-K and M$^\prime$-$\Gamma$-M) and constant energy contour ($E_\mathrm{VBM} = -0.25$~eV.) (d) 2D cut through the 4D ARPES intensity $I(E,k_x,k_y,\theta)$: at different polarization-axis angles ($\theta$), at $E - E_\mathrm{VBM} = -0.25$~eV and for given K and K' valleys.}
\label{fig:sketch}
\end{figure}

\subsection{Polarization-Modulated Angle-Resolved Photoemission Spectroscopy}

The energy and momentum dependence of the photoemission transition dipole matrix element contains rich information on the electronic structure of crystalline solids. However, as in any standard intensity measurements, the phase information is lost, which renders a reconstruction of their Bloch wavefunction challenging. We tackle this challenge by increasing the dimensionality of the measurement: photoemission intensity is recorded while continuously varying the polarization axis direction of linearly polarized XUV ionizing radiation (characterized by the angle $\theta$). By looking at the energy- and momentum-resolved modulation of the photoemission intensity upon polarization rotation, we can access the orientation of hybridized orbitals involved in the photoemission process, which is sensitive to the orbitals' relative phase information. 

To this end, we use our angle-resolved photoemission spectroscopy setup featuring a home-built high-repetition-rate (500 kHz) femtosecond XUV source (polarization-tunable) coupled to a time-of-flight momentum microscope~\cite{Puppin19,maklar20} (see Fig.~\ref{fig:sketch}(a) and Appendix~\ref{app:expdetails}). Measuring the photoemission intensity resolved in energy ($E$) and both parallel momenta ($k_x, k_y$) for each polarization axis direction ($\theta$) yields four-dimensional (4D) data sets $I(E, k_x, k_y, \theta)$. The out-of-plane component $k_\perp$ of the photoelectron momentum vector $\vec{p} = (k_x,k_y,k_\perp)$ is determined by the kinetic energy $E$.

While these multidimensional photoemission data naturally include linear dichroism, the photoemission intensity modulation upon continuous rotation of $\theta$, gives qualitatively new information about the participating orbitals, as detailed below.

\begin{figure}[t]
\centering\includegraphics[width=\columnwidth]{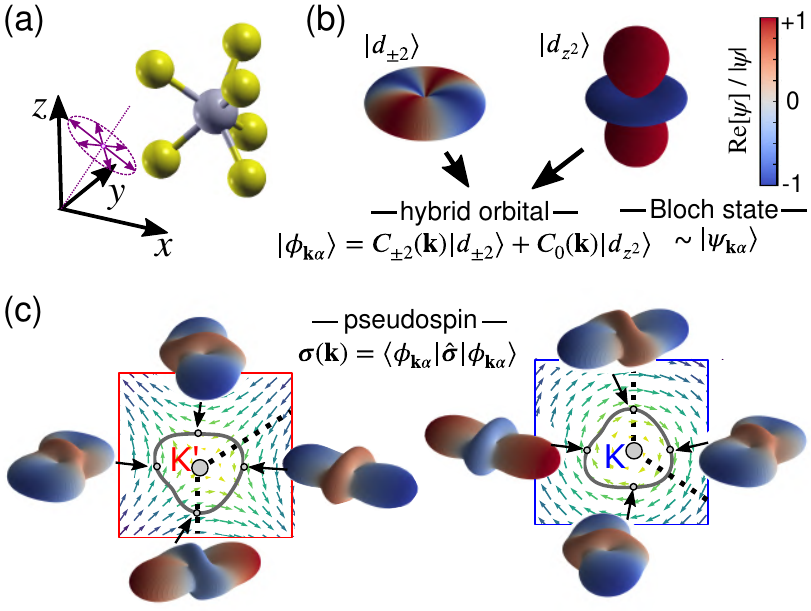}
\caption{\textbf{Wavefunction properties near the valence band maximum, around K/K$^\prime$ valleys:} (a) Sketch of the crystal cell of a monolayer WSe$_2$ and the corresponding coordinate system (used for all three-dimensional plots in this figure). The purple arrows indicate the various polarization directions. (b) Relevant orbitals close to K/K$^\prime$, represented by a constant-value surface of the absolute value, while the color-coding indicates the real part. The Bloch state $|\psi_{\vec{k}\alpha}\rangle$ in the crystal cell in (a) is well approximated by a superposition of the $|d_{\pm 2}\rangle$ and $| d_{z^2}\rangle$ orbitals, which defines the hybrid orbital $|\phi_{\vec{k}\alpha}\rangle$. (c) Plots of the hybrid orbital of
the top valence band at selected momentum points close to the K and K$^\prime$ (corresponding to the boxes in Fig.~\ref{fig:sketch}(b)) valley. There is a one-to-one map of the complex wavefunction coefficients $C_{0,\pm 2}(\vec{k})$ forming the hybrid orbital and the orbital pseudospin $\gvec{\sigma}(\vec{k})$; the corresponding texture is represented by the vector field. The gray thick is a contour of maximum photoemission intensity for typical binding energy.}
\label{fig:hybrid}
\end{figure}

\begin{figure*}[t]
\centering\includegraphics[width=0.9\textwidth]{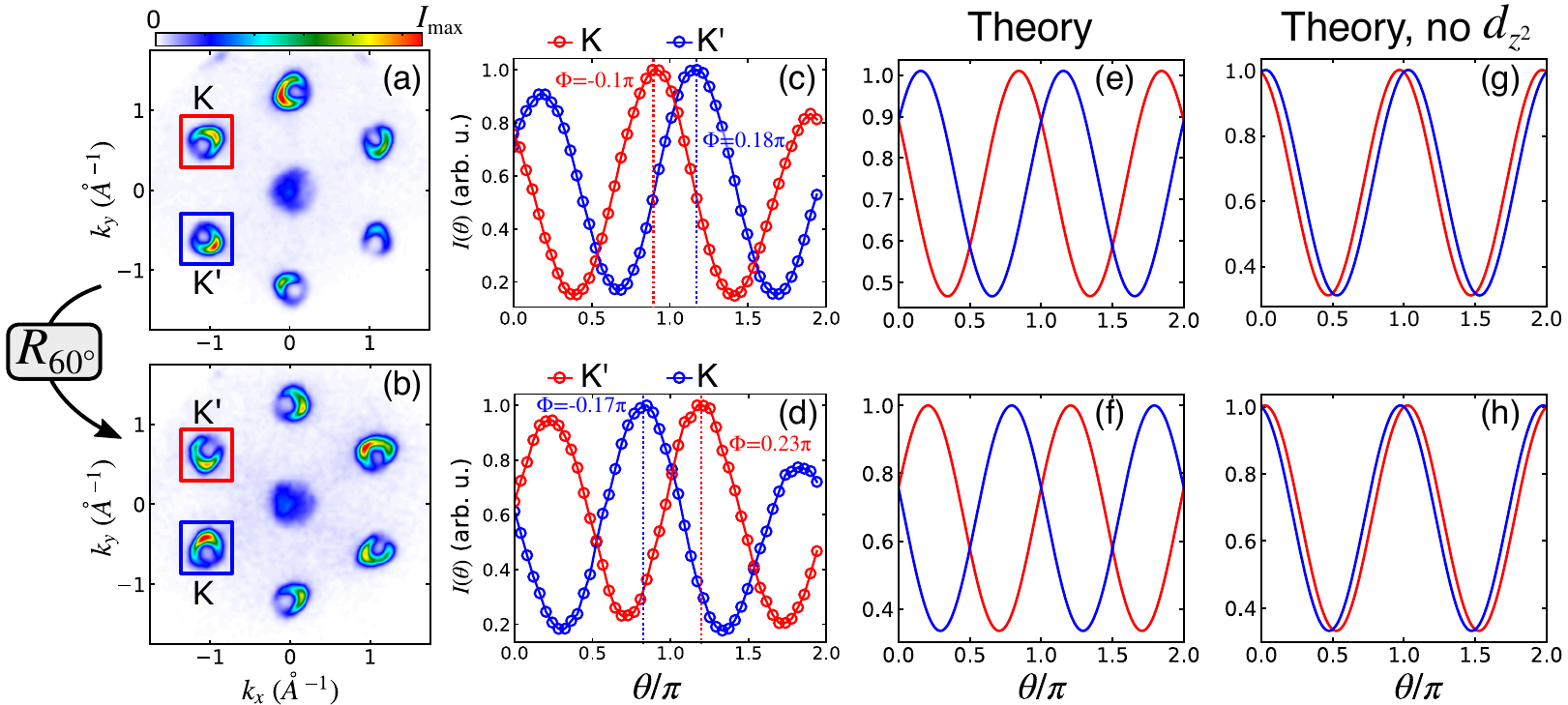}
\caption{\textbf{Valley-resolved polarization-modulated photoemission:} (a)-(b) Constant energy contours (binding energy $E-E_\mathrm{VBM} = -0.25$~eV) for both crystals orientations (averaged over the polarization angle $\theta$). (c)-(d): Valley-dependent polarization-modulated photoemission signal, integrated over the square boxes shown (a),(b). The $\theta$ dependence follows the generic form~\eqref{eq:modulation}; the phase shift $\Phi$ is illustrated by the dotted lines and the value of $\Phi$ for the K/K$^\prime$ valleys is given in the corresponding color. (e)-(f): Calculated polarization-angle modulation of the intensity (analogous to (c)-(d)). (g)-(h): Analogous to (e)-(f), but excluding the $|d_{z^2}\rangle$ orbital contribution. The direction of the light incidence is in the x-z plane.}
\label{fig:theta_dep}
\end{figure*}

The photoemission processes can be described by Fermi's golden rule,
\begin{align}
	\label{eq:fermi}
	I(E,\vec{k},\theta) \propto \left|\langle \vec{k}, E | \vec{e}(\theta)\cdot \hat{\vec{r}} | \psi_{\vec{k}\alpha}\rangle \right|^2 \delta(\en_{\vec{k}\alpha} + \hbar \omega - E) \ ,
\end{align}
where $|\psi_{\vec{k}\alpha} \rangle$ is the initial Bloch state with energy $\en_{\vec{k}\alpha}$, $\vec{e}(\theta)$ the polarization vector of the photons (energy $\hbar \omega$), $\hat{\vec{r}}$ the dipole (or position) operator, and $|\vec{k},E\rangle$ the final states. 
At fixed in-plane momentum $\vec{k}=(k_x,k_y)$ and photoelectron energy $E$, the magnitude of the photoemission intensity is fully determined by the dipole matrix element $M(E,\vec{k},\theta) = \langle \vec{k}, E | \vec{e}(\theta)\cdot \hat{\vec{r}} | \psi_{\vec{k}\alpha}\rangle$ for the band index $\alpha$. 
Note that even if the dipole operator $\hat{\vec{r}}$ itself is ill-defined in periodic systems, the matrix element $M(E,\vec{k},\theta)$ can be rigorously defined in terms of the Berry connection~\cite{Resta_Quantum-Mechanical_1998,Bianco_Mapping_2011}.
In general, the matrix element, and thus the photoemission intensity, is governed by (i) the direction of the outgoing photoelectron $\vec{p} = (k_x,k_y,k_\perp)$, (ii) the light polarization $\vec{e}(\theta)$, and (iii) the orbital character and orientation of the initial state.  In particular, the relative orientation of $\vec{e}(\theta)$ and $\vec{p}$ matters: $I(E,\vec{k},\theta)$ is generally enhanced if they are parallel, and reduced if $\vec{e}(\theta)$ and $\vec{p}$ are orthogonal. 
In our experimental setup, the out-of-plane component is determined by $E = k^2_\perp / 2 + \vec{k}^2/2$ (we use atomic units unless stated otherwise). Hence, $\vec{p}$ is fixed when investigating the signal originating from a particular region in the Brillouin zone, and (at fixed photon energy) the polarization $\vec{e}(\theta)$ is the only remaining external knob to turn to try extracting information on the initial state. 
For the geometry shown in Fig.~\ref{fig:sketch}(a), the $\theta$ dependence of the signal can be discerned by projecting along $s$ (in the $y$ plane) and $p$ (in the $x$-$z$ plane). The polarization vector is then decomposed as $\vec{e}(\theta) = \cos\theta \vec{e}_p + \sin\theta \vec{e}_s$. Introducing the matrix elements with respect to the $s$ ($p$) polarization $M_{s}(E,\vec{k})$ ($M_{p}(E,\vec{k})$) by inserting the corresponding unit vector $\vec{e}_s$ ($\vec{e}_p$), the relevant photoemission matrix element becomes 
$M(E,\vec{k},\theta) = \cos\theta M_p(E,\vec{k}) + \sin\theta M_s(E,\vec{k})$. Inserting into Fermi's golden rule~\eqref{eq:fermi} yields the general form
\begin{align}
  \label{eq:modulation}
  I(E,\vec{k},\theta) = I_0(E,\vec{k}) + B(E,\vec{k}) \cos[2\theta-\Phi(E,\vec{k})] \ .
\end{align}

The photoemission yield modulation upon varying $\theta$ summarized in Eq.~\eqref{eq:modulation} is generic for any system; however, the angle $\theta$ where the intensity is maximized -- determined by the \emph{phase} $\Phi(E,\vec{k})$ -- is extraordinarily sensitive to the initial Bloch state. As further detailed in Appendix~\ref{app:modulation}, $\Phi(E,\vec{k})$ depends on the magnitude of the photoemission matrix elements with respect to $s$ and $p$ polarized light \emph{and} their phase relation, which encodes the phase of the underlying orbitals and interference effects.

\subsection{Orbital Character and Photoemission Matrix Elements}

To connect the orbital character of the Bloch state $|\psi_{\vec{k}\alpha}\rangle$ to the photoemission signal, it is useful to introduce the Wannier representation~\cite{marzari_maximally_2012} 
\begin{align}
  \label{eq:wannier}
  \psi_{\vec{k}\alpha}(\vec{r}) &= \frac{1}{N}\sum_{\vec{R},m} e^{i \vec{k}\cdot\vec{R}} C_{m\alpha}(\vec{k}) w_m(\vec{r}-\vec{R})
  \nonumber \\
   &\equiv \frac{1}{N}\sum_{\vec{R}} e^{i \vec{k}\cdot\vec{R}} \phi_{\vec{k}\alpha}(\vec{r}-\vec{R}) \ ,
\end{align}
where $\vec{R}$ labels all $N$ unit cells, while $w_m(\vec{r})$ are the Wannier functions that can be paralleled to atomic orbitals~\footnote{We are using a different normalization of the Bloch wavefunction compared to the commonly used convention from ref.~\cite{marzari_maximally_2012}, such that the normalization factor cancels out when computing photoemission matrix elements}.
The coefficients $C_{m\alpha}(\vec{k})$ connect orbital ($m$) and band ($\alpha$) space, thus determining the $\vec{k}$-dependent superposition of the orbitals. This hybridization is conveniently captured by introducing the hybrid orbital $\phi_{\vec{k}\alpha}(\vec{r}) = \sum_{m} C_{m\alpha}(\vec{k}) w_m(\vec{r})$. The hybrid orbital provides a useful way of understanding the Bloch wavefunction: for fixed $\vec{k}$, the Bloch state is constructed by a periodic replica of $\phi_{\vec{k}\alpha}(\vec{r})$ with a phase factor $e^{i \vec{k}\cdot \vec{R}}$. In the common situation where the Wannier functions $w_m(\vec{r})$ are sufficiently localized and contributions from neighboring unit cells can be neglected, the hybrid orbital is a good approximation (up to a normalization factor) to the actual Bloch state for $\vec{r}$ within a selected unit cell. The concept of the hybrid orbital also connects directly to the photoemission matrix element (see Appendix~\ref{app:tbmel}). In case the relevant Wannier orbitals $w_m(\vec{r})$ are localized on a single site, one can show
\begin{align}
  \label{eq:mel_hybrid}
  M(E,\vec{k},\theta) = \langle \vec{k},E | \vec{e}(\theta) \cdot \hat{\vec{r}}| \phi_{\vec{k}\alpha}\rangle \ .
\end{align}
Hence, the photoemission matrix element can be understood analogously to \emph{atomic} photoemission, where the atomic orbitals are replaced by the hybrid orbitals.

Specifically for $2H$-WSe$_2$, the quantum nature of the top valence band is dominated by
$|d_{z^2}\rangle$, $|d_{x^2-y^2}\rangle$ and $|d_{xy}\rangle$ orbitals localized at the W atoms~\cite{fang_ab_2015}. 
Near the valence band maximum (VBM) the hybrid orbital is given by $|\phi^{\mathrm{K,K}^\prime}_{\vec{k}}\rangle \approx [C_{\pm}(\vec{k}) |d_{\pm 2}\rangle + C_0(\vec{k}) | d_{z^2}\rangle] \otimes |\uparrow, \downarrow\rangle$, where $|d_{\pm 2}\rangle = [|d_{x^2-y^2}\rangle \pm i |d_{xy}\rangle]/\sqrt{2}$ are magnetic orbitals~\footnote{We suppress the band index $\alpha$ in what follows as we focus on the top valence band.}. For coordinates 
within the crystal cell (including the nearest-neighbor Se atoms) displayed in Fig.~\ref{fig:hybrid}(a), the Wannier orbitals are sketched in Fig.~\ref{fig:hybrid}(b). 
Even though the out-of-plane $|d_{z^2}\rangle$ orbital contribution vanishes at exactly $\vec{k} =$ K, K$^\prime$, the $\vec{k}$-dependent interference between these orbitals has profound impact when moving slightly away from $\vec{k} =$ K, K$^\prime$. This interference manifests in the $\vec{k}$-dependent spatial orientation of the hybrid orbital, as illustrated in Fig.~\ref{fig:hybrid}(c) where we show $\phi^{\mathrm{K,K}^\prime}_{\vec{k}}(\vec{r})$ for different $\vec{k}$-points close to the Dirac valleys K/K$^\prime$. The hybrid orbital is reminiscent of $|d_{z^2}\rangle$ orbital rotated to the $x$--$y$ plane, and its orientation exhibits a pronounced momentum dependence.

This orbital texture in momentum space is closely related to the concept of orbital pseudospin $\sigma^{\mathrm{K,K}^\prime}_\nu(\vec{k}) = \langle \psi^{\mathrm{K,K}^\prime}_{\vec{k}} | \hat{\sigma}_\nu | \psi^{\mathrm{K,K}^\prime}_{\vec{k}} \rangle = \langle \phi^{\mathrm{K,K}^\prime}_{\vec{k}} | \hat{\sigma}_\nu | \phi^{\mathrm{K,K}^\prime}_{\vec{k}} \rangle$ ($\hat{\sigma}_\nu$ denote the Pauli matrices, $\nu=x,y,z$). There is a one-to-one correspondence between the complex coefficients $C_{0,\pm 2}(\vec{k})$ and $\sigma_\nu(\vec{k})$; the pseudospin texture is an elegant way to visualize the complex coefficients. The in-plane pseudospin texture is shown in Fig.~\ref{fig:hybrid}(c). In general, orbital pseudospin relates to the Berry curvature and topological properties of materials~\cite{qi_topological_2008}. For 2H-WSe$_2$, the $\sigma^{\mathrm{K,K}^\prime}_z(\vec{k})$ captures the weight of the $|d_{\pm 2}\rangle$ and $|d_{z^2}\rangle$ orbital, respectively, while the in-plane texture encodes interference. The pseudospin texture of 2H-WSe$_2$ manifests itself in the characteristic momentum dependence of the photoemission signal within the K/K$^\prime$ valleys~\cite{beaulieu_revealing_2020-1}.

Now, we discuss the experimentally measured as well as the calculated modulation of the photoemission intensity upon varying the angle $\theta$ (see Fig.~\ref{fig:sketch}(a)). For all calculations presented in this work, we employ the tight-binding (TB) model for a monolayer WSe$_2$ from ref.~\cite{liu_three-band_2013} (details are presented in Appendix~\ref{app:tbdetails}). We benchmarked the model against a first-principle model obtained from computing projective Wannier functions~\cite{schuler_gauge_2021} and found that the orbital character is accurately reproduced by the TB model for a moderate region around the valleys. The model includes the $|d_{z^2}\rangle$ and $|d_{\pm 2}\rangle$ orbitals only; however, the full crystal symmetry is incorporated into the model, thus also capturing the effective hybridization of W and Se atoms (see supplemental materials~\cite{supplement} for a discussion). The model is combined with the plane-wave approximation to the final states~\footnote{For photon energies in the XUV regime, and this specific system, we have ensured that final state effects play only a minor role (see Appendix~\ref{app:tbmel}).}. The layered structure of bulk $2H$-WSe$_2$ gives rise to intra- and interlayer hybridization influencing the orbital and spin character of the bands in the vicinity of the K, K$^\prime$ valley~\cite{rostami_layer_2019,razzoli_selective_2017}. However, interlayer hybridization leads to only a small correction to the orbital character~\cite{fang_ab_2015}; furthermore, layer-resolved first-principle ARPES calculations from ref.~\cite{beaulieu_revealing_2020-1} have shown that photoemission signal can be attributed almost solely to the topmost layer. Therefore, the (spin-integrated) photoemission signal can be understood in terms of monolayer WSe$_2$. 

The experimentally measured photoemission intensity modulation contrast upon varying $\theta$ is very pronounced (Fig.~\ref{fig:theta_dep}(a)--(d)). The valley-integrated photoemission intensity $I_\mathrm{int}(E,\theta)$ ($\vec{k}$-integrated over the boxes shown Fig.~\ref{fig:theta_dep}(a),(b)) shows an intrinsic phase shift between K and K$^\prime$ valleys. To confirm that this phase shift is an intrinsic property of the crystals and does not originate from spurious experimental geometry effects, we rotated the crystal by $\mathrm{60^{\circ}}$, acting as an effective in-plane time-reversal transformation, \textit{i.\,e.} $\mathrm{K} \leftrightarrow \mathrm{K}^\prime$~\cite{fang_ab_2015}. Upon effective time-reversal transformation (swapping the valley indexes), the relative phase shift changes sign, indicating that the polarization-modulated photoemission yield is sensitive to intrinsic valley-resolved properties of the crystal. We have calculated the photoemission intensity for varying $\theta$ using Eq.~\eqref{eq:fermi} from the TB model and analyzed in the same way as the experimental data.
Apart from the absolute scale, the angular dependence of the photoemission intensity and the phase shifts are well reproduced by our theoretical calculations (Fig.~\ref{fig:theta_dep}(e)--(f)).
Excluding the $|d_{z^2}\rangle$ orbitals strongly diminish the intrinsic phase shift between K and K$^\prime$ valleys (Fig.~\ref{fig:theta_dep}(g)--(h)), thus underlining the interplay of $|d_{z^2}\rangle$ and $|d_{\pm 2}\rangle$ orbitals in the emergence of the experimentally observed polarization-modulated photoemission signal. This conclusion is further underpinned by inspecting the intensity modulation if we assume the orbital character of the top valence band is purely given by $|d_{z^2}\rangle$ or $|d_{\pm 2}\rangle$ (see Appendix~\ref{app:atomiclim}). In this atomic limit, the phase of the intensity modulation does not match the experiments; especially the change upon K$\leftrightarrow$K$^\prime$ exhibits the wrong behavior, while rotation by 60$^\circ$ does not show any effect.

\subsection{Energy- and Momentum-Resolved Fourier Analysis}

In an attempt at obtaining deeper insights about the link between our new measurement methodology and the electronic properties of the crystal, we have performed a fully energy- and momentum-resolved Fourier transform analysis along the XUV polarization axis in order to extract oscillation amplitude and phase of the signal in specific energy-momentum region of the electronic structure. The Fourier-transformed signal
\begin{align}
	\label{eq:fourier_int}
	I_m(E,k_x,k_y) = \int^{2\pi}_0\frac{d\theta}{2\pi}e^{i m \theta} I(E,k_x,k_y,\theta)
\end{align}
is only nonzero for $m=0,\pm 2$, as the intensity~\eqref{eq:modulation} is dependent on $2\theta$. While $m=0$ corresponds to the $\theta$-averaged intensity (which is identical to $I_0(E,\vec{k})$ in Eq.~\eqref{eq:modulation}), $I_2(E,k_x,k_y)$ is a \emph{complex} quantity encoding information about the amplitude/real and phase/imaginary information of the photoemission modulation upon rotating the polarization axis of the XUV. For the experimental geometry (Fig.~\ref{fig:sketch}(a)), direct evaluation yields
\begin{align}
	\label{eq:i2_real}
	\mathrm{Re}[I_2(E,\vec{k})] &= \frac14 \left[|M_s(E,\vec{k})|^2 - |M_p(E,\vec{k})|^2\right] g(E,\vec{k}) \\  &= \frac14 I_\mathrm{LDAD}(E, \vec{k}) \ , \nonumber \\
	\label{eq:i2_imag}
	\mathrm{Im}[I_2(E,\vec{k})] &= -\frac12 \mathrm{Re}\left[(M_s(E,\vec{k}))^* M_p(E,\vec{k})\right] g(E,\vec{k})  \ , 
\end{align}
where $M_s(E,\vec{k})$ and $M_p(E,\vec{k})$ denote the matrix elements with respect to $s$- or $p$- polarized light and where $g(E,\vec{k})=\delta(\en_{\vec{k}\alpha} + \hbar{\omega} - E)$. While the real part~\eqref{eq:i2_real} contains information on the linear dichroism of the photoemission intensity with respect to $s$- or $p$- polarized light (equivalent to the LDAD), the imaginary part~\eqref{eq:i2_imag} captures \emph{interference} between these channels. We stress that the latter is a new quantity that cannot be obtained by solely measuring the photoemission intensity using $s$- or $p$-polarized photons. This interferometric quantity, revealing the relative phase between $M_s(E,\vec{k})$ and $M_p(E,\vec{k})$, is available within the context of our novel polarization-modulated ARPES approach~\footnote{In principle, measuring the photoemission signal with respect to three orthogonal polarization directions would yield similar information. However, this would require changing the angle of incidence, which is very hard to accomplish in most ARPES setups.}.
This relative phase is also fundamentally connected to the circular dichroism in photoelectron angular distributions (CDAD), as discussed below.

\begin{figure}[t]
\centering\includegraphics[width=\columnwidth]{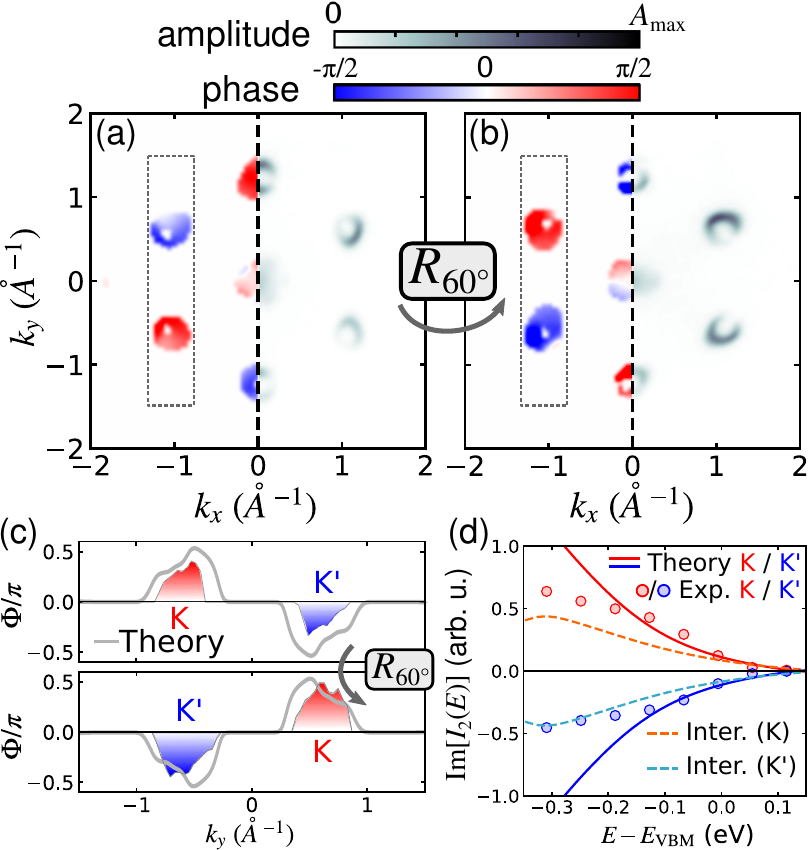}
\caption{\textbf{Energy- and momentum-resolved Fourier analysis of the polarization-modulated ARPES signals:} (a)--(b) Fourier amplitude $|I_2(E,\vec{k})|$ (white-to-black colormap, right sub-panel) and phase $\Phi(E,\vec{k})$ (blue-to-red colormap, left sub-panel) of the photoemission modulation, for $E - E_\mathrm{VBM}$ = -0.25~eV. (c) Averaged phase (along a vertical cut and integrated over $k_x$ as indicated by the dashed box in (a), (b)), extracted from both the experimental data in (a)--(b) and the theory. (d) Valley-integrated imaginary part $\mathrm{Im}[I_2(E)]$ of the Fourier amplitude~\eqref{eq:fourier_int}, comparing experiment and theory at K (red) and K$^\prime$ (blue), respectively. The dashed lines represent the corresponding interference contribution $\mathrm{Im}[I^\mathrm{int}_2(E)]$.}
\label{fig:phase_im}
\end{figure}{}

The energy- and momentum-resolved phase $\Phi(E,\vec{k})=\mathrm{arg}[I_2(E,\vec{k})]$ (which is identical to the phase shift in Eq.~\eqref{eq:modulation}) is presented in Fig.~\ref{fig:phase_im}(a)--(b), for a binding energy of $E - E_\mathrm{VBM}\sim -0.25$~eV. We observe a phase sign-flip for adjacent valleys, as well as a phase sign reversal upon effective time-reversal operation ($\mathrm{60^{\circ}}$ crystal rotation). This is a clear indication that the distinct orbital character of the Bloch state -- which exhibits a distinct texture at K or K$^\prime$, respectively -- is responsible for the observed sign change. Fig.~\ref{fig:phase_im}(c) shows the phase integrated along $k_x$ (going from K to K$^\prime$, and vice-versa), which is well captured by our TB model calculations.

Inspecting the real~\eqref{eq:i2_real} and imaginary part~\eqref{eq:i2_imag} of the Fourier signal (see supplemental materials~\cite{supplement}), we notice that the sign of the phase $\Phi(E,\vec{k})$ and $\mathrm{Im}[I_2(E,\vec{k})]$ are qualitatively identical. Comparing the imaginary part to the theoretical results is straightforward. The theory allows for decomposing the valley-integrated signal (around the K / K$^\prime$ valleys in Fig.~\ref{fig:phase_im}(a)) into $\mathrm{Im}[I_2(E)] = \mathrm{Im}[I^\mathrm{inc}_2(E)] + \mathrm{Im}[I^\mathrm{int}_2(E)]$,
where the $\mathrm{Im}[I^\mathrm{inc}_2(E)] = \mathrm{Im}[I^\mathrm{z^2}_2(E)] + \mathrm{Im}[I^\mathrm{\pm 2}_2(E)]$ by incoherently adding signal originating from only the $|d_{z^2}\rangle$ or $|d_{\pm 2}\rangle$ orbital, while $\mathrm{Im}[I^\mathrm{int}_2(E)] = \mathrm{Im}[I_2(E)] - \mathrm{Im}[I^\mathrm{inc}_2(E)]$ denotes the remaining interference contribution. This analysis underpins that the interference of the $|d_{z^2}\rangle$ and $|d_{\pm 2}\rangle$ orbitals is the predominant contribution close to the VBM. 

In contrast to what might be expected from the pseudospin texture (Fig.~\ref{fig:hybrid} (c)), the phase $\Phi(E,\vec{k})$ exhibits almost no momentum dependence around the K/K$^\prime$ valleys (intravalley). To understand the quantity $\Phi(E,\vec{k})$ better, we have derived the relation to the photoemission matrix elements $M_s(E,\vec{k})$, $M_p(E,\vec{k})$ explicitly (see Appendix~\ref{app:modulation}). 
In brief, the phase $\Phi(E,\vec{k})$ is determined by many factors: (i) the phase difference $\phi_p - \phi_s$, where $M_{s,p}(E,\vec{k}) = |M_{s,p}(E,\vec{k})| e^{i \phi_{s,p}(E,\vec{k})}$ is the phase of the matrix element itself, and (ii) the ratio $|M_{s}(E,\vec{k})|/|M_{p}(E,\vec{k})|$. The momentum-dependent hybridization represented by the pseudospin texture manifests in the orientation and the phase of the hybrid orbital (illustrated in Fig.~\ref{fig:hybrid}(c)) and thus also in $M_{s,p}(E,\vec{k})$. However, this momentum dependent phase mostly cancels out when taking the phase difference $\phi_p(E,\vec{k}) - \phi_s(E,\vec{k})$. Hence, the phase $\Phi(E,\vec{k})$ displays only a weak intravalley momentum dependence. 

Nevertheless, information on the wavefunction is encoded in $\Phi(E,\vec{k})$: the sign changes with respect to adjacent valleys are consistent with the magnetic orbital character $|d_{\pm 2}\rangle$, as evidenced by rotating the crystal by 60$^\circ$. Additional calculations (see supplemental materials~\cite{supplement}) qualitatively yield the same picture. We interpret this behavior as an interplay of interference of $|d_{z^{2\rangle}}$ and $|d_{\pm 2}\rangle$, their spatial orientation, and the geometry of our experimental setup. All of these factors play a role: the alternating sign of $\Phi(E,\vec{k})$ with respect to adjacent valleys is suppressed if (i) either the $|d_{z^2}\rangle$ or $|d_{\pm 2}\rangle$ orbital is excluded, (ii) if the interference of these orbitals is switched off, (iii) at larger photon energy (see supplemental materials~\cite{supplement}). 

\begin{figure*}[t]
\centering\includegraphics[width=0.9\textwidth]{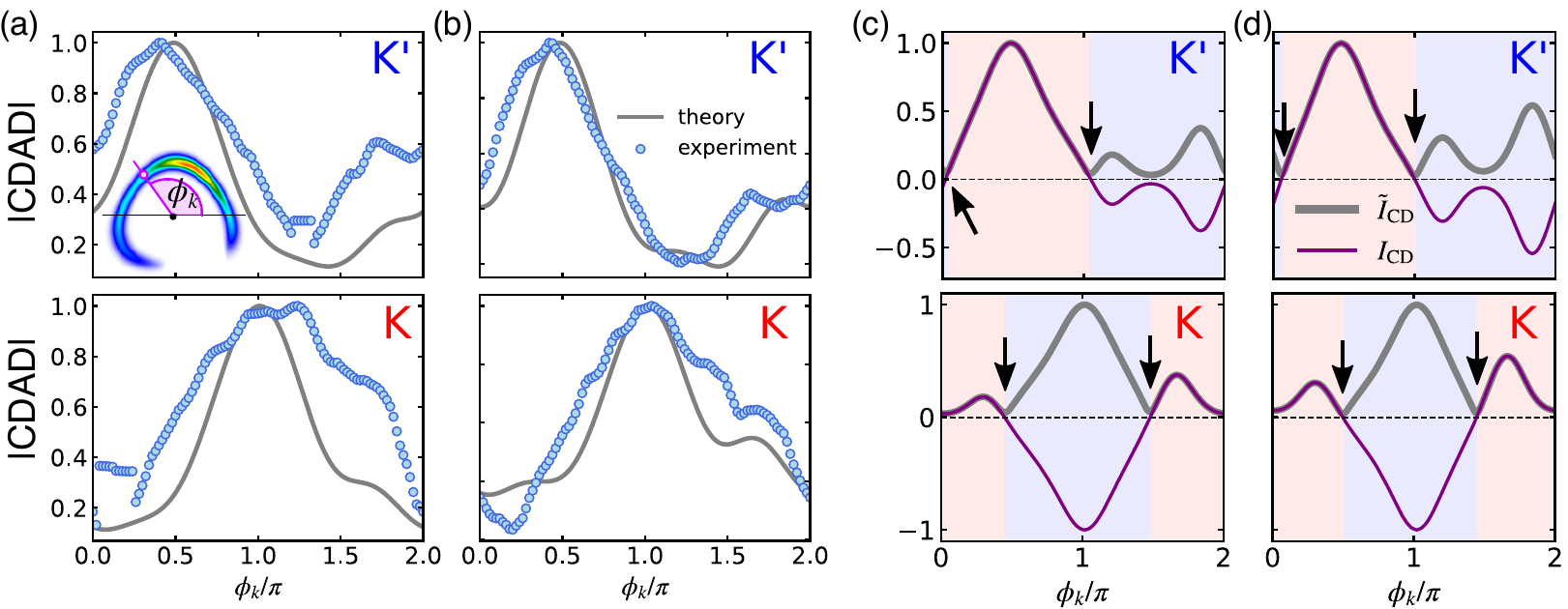}
\caption{\textbf{Circular dichroism without circular photons:} (a), (b): Absolute value of the CDAD extracted from the experimental data and theory via Eq.~\eqref{eq:cdad_rc} at fixed binding energy $E - E_\mathrm{VBM}=-0.18$~eV (a) and $E - E_\mathrm{VBM}=-0.25$~eV (b), as function the angle $\phi_k$ tracing the intensity. The inset illustrates how the angle $\phi_k$ is measured along the contour of maximum intensity. (c),(d): Theoretical reconstructed ($\widetilde{I}_\mathrm{CD}(E,\vec{k})$) and calculated CDAD ($I_\mathrm{CD}(E,\vec{k})$) as in (a), (b). The arrows indicate the kink positions that can be used to determine sign changes (indicated by shaded background).}
\label{fig:four2cd}
\end{figure*}

\subsection{Circular Dichroism without Circular Photons}

On a more fundamental level, the imaginary part~\eqref{eq:i2_imag} of the polarization-modulated photoemission, that we refer to as \textit{Fourier Dichroism in Photoelectron Angular Distributions} (FDAD), is the missing piece to measuring the phase of the complex dipole matrix elements directly, if Circular Dichroism in Photoelectron Angular Distributions (CDAD) is additionally available. Indeed, let us consider the experimental scheme as in Fig.~\ref{fig:sketch}(a), but using left-handed circularly polarized (LCP) or right-handed circularly polarized (RCP) light. The CDAD is then defined by $I_\mathrm{CD}(E,\vec{k}) = I_\mathrm{LCP}(E,\vec{k}) - I_\mathrm{RCP}(E,\vec{k})$. Substituting the corresponding polarization vector $\vec{e}_\mathrm{RCP/LCP}$ into Eq.~\eqref{eq:fermi}, one obtains
\begin{align}
	\label{eq:cdad}
	I_\mathrm{CD}(E,\vec{k}) =-2 \mathrm{Im}\left[(M_s(E,\vec{k}))^* M_p(E,\vec{k})\right] g(E,\vec{k}) \ .
\end{align}
Comparing Eq.~\eqref{eq:cdad} to Eq.~\eqref{eq:i2_imag} we notice a striking similarity: instead of the imaginary part of the complex quantity 
\begin{align*}
  Z(E,\vec{k}) &\equiv (M_s(E,\vec{k}))^* M_p(E,\vec{k}) \\ &
  = |Z(E,\vec{k})|e^{i (\phi_p(E,\vec{k}) - \phi_s(E,\vec{k}) )} \ ,
\end{align*}
the imaginary part of the Fourier signal~\eqref{eq:i2_imag} - FDAD - provides access to the real part of $Z(E,\vec{k})$. While $|Z(E,\vec{k})|$ can be extracted by measuring the photoemission intensity for  $s-$ and $p-$ polarized light separately, the relative phase $\phi_p - \phi_s$ is available by combining FDAD \emph{and} the CDAD. Since the global phase of $M_s(E,\vec{k})$ or $M_p(E,\vec{k})$ is not relevant (it does not manifest in any observable), obtaining $\phi_p - \phi_s$ allows extracting complete information on the complex matrix elements $M_{s/p}(E,\vec{k})$. 
Because the CDAD and FDAD are complementary parts of the same complex quantity, they are fundamentally linked. In fact, this intricate relationship can be exploited to obtain insights into the CDAD without using circularly polarized XUV light. Using Eq.~\eqref{eq:i2_imag}, this link can be expressed as
\begin{align}
	\label{eq:cdad_rc}
	\widetilde{I}_\mathrm{CD}(E,\vec{k}) = \sqrt{I_s(E,\vec{k}) I_p(E,\vec{k}) - 4 \mathrm{Im}[I_2(E,\vec{k})]} \ .
\end{align}
Note that only the absolute value of the CDAD can be extracted due to a missing absolute phase information to directly link a continuous scan of the linear XUV polarization axis to circularly polarized light (indicated by the tilde in Eq.~\eqref{eq:cdad_rc}, $|I_\mathrm{CD}(E,\vec{k})| = \widetilde{I}_\mathrm{CD}(E,\vec{k})$). Nevertheless, fine details on the momentum dependence of the circular dichroism can still be extracted as demonstrated in Fig.~\ref{fig:four2cd}(a)--(b). 

To this end we substituted the imaginary part $\mathrm{Im}[I_2(E,\vec{k})]$ (obtained by Fourier transforming the experimental data via Eq.~\eqref{eq:fourier_int}) into Eq.~\eqref{eq:cdad_rc}. Note that the valley-averaged CDAD provides a direct map of the Berry curvature of WSe$_2$, as demonstrated by previous experiments~\cite{cho_experimental_2018,cho_studying_2021} and theory~\cite{schuler_local_2020-1}. However, the CDAD exhibits a fine structure even within a single valley (which depends on the experimental geometry). Thus, for a given binding energy, we show the extracted CDAD as a function of the azimuthal angle $\phi_k$, which traces the constant energy contour (see inset in Fig.~\ref{fig:four2cd}(a)). By broadening the momentum distribution of the theoretical data to mimic the experimental momentum resolution (we used Gaussian smearing of $\Delta k\sim 0.05$~\AA$^{-1}$), we find a striking agreement between experiment and theory for both valleys (Fig.~\ref{fig:four2cd}(a)--(b)). This agreement implies that our novel measurement procedures allow getting information about CDAD, without using circularly polarized photons.  

While the sign of the reconstructed CDAD is, in principle, not available, sharp kinks near zero would indicate a sign change. The momentum resolution of the experiment is not sufficient to identify such sharp features; however, the excellent qualitative agreement between theory and experiment for all considered quantities described above allows us to extrapolate to a better resolution
\footnote{In the calculations we replace that Dirac delta function by a Gaussian with energy smearing $\Delta \varepsilon$. For most results presented here, we use $\Delta \varepsilon=2\times 10^{-3}$ a.\,u.; for Fig.~\ref{fig:four2cd}(c)--(d) we used $\Delta \varepsilon=10^{-3}$ a.\,u., which corresponds to a momentum resolution of $\Delta k \approx 0.035$~\AA$^{-1}$.}. 
This scenario is explored in Fig.~\ref{fig:four2cd}(c)--(d), where we compare the calculated CDAD to the reconstructed signal (via Eq.~\eqref{eq:cdad_rc}). Due to taking the absolute value, kinks (indicated by black arrows) appear in the reconstructed CDAD, which allows pinpointing sign changes. Therefore, up to an absolute sign ambiguity for each valley, the full CDAD can be extracted from our experimental data, where no circularly polarized photons were used. Note that the momentum resolution required to identify the kinks is well within current experimental capabilities.

\subsection{Orbital Pseudospin and Bloch Wavefunction Reconstruction}


\begin{figure}[t]
\centering\includegraphics[width=\columnwidth]{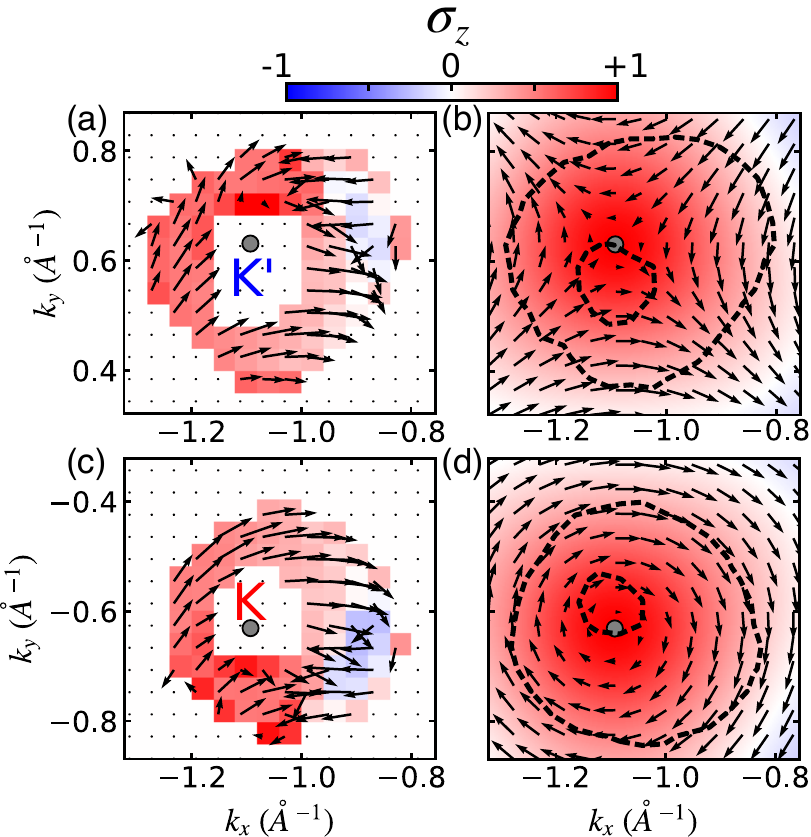}
\caption{\textbf{Reconstruction of orbital pseudospin:} (a), (b): Pseudospin texture reconstructed from experimental (with emulated CDAD) data. The in-plane pseudospin is represented by the arrows, while the $z$-component is indicated by the color map. We present results for $E-E_\mathrm{VBM}=-0.25$~eV, where the reconstruction procedure is most stable. (c), (d): Orbital pseudospin from the TB model. The black dashed lines indicate the region in which the intensity $I_\mathrm{av}(E,\vec{k}) > 0.1 I_\mathrm{max}$ ($I_\mathrm{max}$ is the maximum intensity).   }
\label{fig:pseudo_rc}
\end{figure}

The complementary information encoded in the imaginary part~\eqref{eq:i2_imag} - FDAD - and the circular dichroism can be exploited even further if the CDAD is measured in the same geometry. Our current setup does not allow us to generate circularly polarized XUV photons. However, complementing the experimental data with the CDAD calculated from our theoretical model -- which emulates experimental data -- allows us to showcase which new information could be extracted.
To this end, we have calculated 
\begin{align}
    \label{eq:cdad_emul}
    I^\mathrm{emul}_\mathrm{CD}(E,\vec{k}) = \frac{I^\mathrm{TB}_\mathrm{CD}(E,\vec{k})}{I^\mathrm{TB}_\mathrm{av}(E,\vec{k})} I^\mathrm{exp}_\mathrm{av}(E,\vec{k}) \ ,
\end{align}
where $I_\mathrm{av}(E,\vec{k}) = (I_s(E,\vec{k}) + I_p(E,\vec{k}) )/2$ is the unpolarized intensity. The superscript TB (exp) stands for theoretical (experimental) spectra. 
The TB model includes two orbitals only ($d_{z^2}$ and $d_{\pm 2}$), such that the Bloch state $|\psi^{\mathrm{K},\mathrm{K}^\prime}_{\vec{k}\alpha}\rangle$ is fully characterized by the three components of pseudospin $\sigma^{\mathrm{K,K}^\prime}_\nu(\vec{k})$ ($\nu=1,2,3$). It is convenient to express the photoemission intensity~\eqref{eq:fermi} in terms of $\sigma^{\mathrm{K,K}^\prime}_\nu(\vec{k})$ and the atomic dipole matrix elements $M_{z^2, \pm 2}(E, \vec{k}) = \langle \vec{k},E | \vec{e}\cdot \hat{\vec{r}}  | d_{z^2,\pm 2} \rangle$. 
Approximating the orbitals as the product of a radial wavefunction and a spherical harmonic $Y^2_{0,\pm 2}$ allows to characterize $M_{z^2, \pm 2}(E, \vec{k})$ by a few parameters, which can be fixed by comparing to experimental spectra with $s-$ and $p-$ polarized light, respectively
\footnote{For WSe$_2$, the suppression of the photoemission intensity roughly along the K--M or K--M$^\prime$ direction is a characteristic feature that arises due to the interference of the $d_{z^2}$ and $d_{\pm 2}$ orbitals~\cite{rostami_layer_2019} and closely related to the orbital pseudospin~\cite{beaulieu_revealing_2020-1}. A more detailed discussion can be found in the supplemental materials~\cite{supplement}.}.
Since we also have three independent quantities $\vec{X}(E,\vec{k}) = (I_\mathrm{CD}(E,\vec{k}), \mathrm{Re}[I_2(E,\vec{k})],\mathrm{Im}[I_2(E,\vec{k})] )$ at our disposal, can we use this information to reconstruct the three-dimensional orbital pseudospin? While this methodology is a very ambitious goal at the current stage, the attempt turns out to be instructive.

Based on the atomic matrix elements $M_{z^2, \pm 2}(E, \vec{k})$, we can express the CDAD and the FDAD as linear functions of $\sigma^{\mathrm{K,K}^\prime}_\nu(\vec{k})$, which yields a $3\times 3$ system of equations of the form $\vec{X}(E,\vec{k}) = \vec{A}(E,\vec{k})\gvec{\sigma}^{\mathrm{K,K}^\prime}(\vec{k}) + \vec{B}(E,\vec{k})$, as detailed in Appendix~\ref{app:reconstruct}.  We have solved these equations -- using the calculated matrix elements, the experimental FDAD and the emulated CDAD -- by a least-square minimization while constraining the pseudospin to $\sum_\nu(\sigma^{\mathrm{K,K}^\prime}_\nu(\vec{k}))^2=1$. This is possible for all momenta $\vec{k}$ where the signal at fixed binding energy is large enough (we chose a threshold of 0.1 of the maximum value). The reconstructed pseudospin for the $E-E_\mathrm{VBM}=-0.25$~eV is presented in Fig.~\ref{fig:pseudo_rc}(a)--(b) for K and K$^\prime$, respectively. Scanning through the binding energies allows, in principle, to systematically reconstruct the pseudospin texture in the relevant region in momentum space, albeit the pseudospin picture (i.\,e. where only two orbitals are relevant) breaks down further away from K/K$^\prime$. We have asserted the stability of the reconstruction procedure by retrieving the pseudospin texture from purely theoretical input, which yields exactly the texture from the TB model.

Comparing the reconstructed texture of $\sigma^{\mathrm{K,K}^\prime}_\nu(\vec{k})$ to the calculated one (Fig.~\ref{fig:pseudo_rc}(c)--(d)), we observe some features that are in qualitative agreement. First, the $z$ component (represented by the color-coding) shows some important similarities. Close to the K/K$^\prime$ point, the value of the reconstructed $\sigma_z$ is close to $+1$ (indicating that the $|d_{\pm 2}\rangle$ orbital dominates); moving away from K/K$^\prime$, this value decreases. This behavior is anisotropic -- which is in accordance with the calculated pseudospin $z$ component. The in-plane components of the pseudospin (represented by the black arrows) also show some agreement with the theoretical texture, albeit there are more discrepancies. The saddle-point structure around K$^\prime$ in Fig.~\ref{fig:pseudo_rc}(b) with the texture pointing into the center (out of the center) along the diagonal (orthogonal to the diagonal) is approximately retrieved in the reconstructed texture (Fig.~\ref{fig:pseudo_rc}(a)). Also, parts of the circular winding around K (Fig.~\ref{fig:pseudo_rc}(c)--(d)) are recovered in the upper half.


Deviations are mostly attributed to two factors: (i) the momentum and energy resolution, and (ii) limited predictive power of the photoemission model. Systematic theoretical improvements of the treatment of the photoemission matrix elements -- and thus the retrieval of $\sigma^{\mathrm{K,K}^\prime}_\nu(\vec{k})$ -- beyond the presented model can be achieved by more accurate calculation of the final states and by taking into account non-spherical deformations of the Wannier orbitals. In some cases, such corrections can be constructed from the crystal symmetry in terms of a ligand field theory~\cite{daul_ligand_2013}; in general, systematic corrections of the orbitals introduce additional parameters to be determined by fitting to characteristic experimental signatures.

We stress that in principle, besides the atomic matrix elements, no further input from theory is required (if it can ensure that only two orbitals are relevant). In particular, the band structure beyond the experimentally obtained intensity does not enter the reconstruction procedure.
Hence, a practical route for applying the procedure to other systems is to fit the atomic matrix elements to specific features in equilibrium (for instance, our atomistic model for WSe$_2$ reproduces the dark corridor). The obtained matrix elements are the only required theoretical input for tracing the impact of light-dressing, coherent excitation, or strain onto the pseudospin texture. Since the modeling of the atomic matrix elements is generic, the presented reconstruction procedure can be applied to many systems where only two relevant orbitals are involved.

\section{Discussion/Conclusion}

We introduced the continuous rotation of the polarization axis as a new measurement methodology in ARPES, which adds a new dimension allowing to define a genuine new observable. In particular, the intensity modulation upon varying the polarization angle is generic, and the corresponding phase of the modulation is related to the phase of the photoemission matrix elements and thus to the electronic wavefunction of the system. Taking the Fourier transform of the periodically modulated photoemission yield leads to the definition of a quantity that we introduce as \textit{Fourier dichroism in the photoelectron angular distribution - FDAD}, which is an interferometric quantity complementary to linear and circular dichroism.


Exploiting the fundamental link of FDAD and circular dichroism (CDAD) allows for extracting the absolute value of CDAD without using circularly polarized XUV pulses. This is a major advance since the table-top generation of circularly polarized XUV is challenging and its combination with ARPES endstation has not been reported yet. The extension of our approach to time-resolved CDAD experiments without the need for circular XUV photons
is conceptually and practically straightforward, as the present experimental setup is already operating with femtosecond pulses. Hence, tracking the time-resolved FDAD will allow tracking ultrafast light-induced topological phase transition characterized by creation or annihilation of local Berry curvature, for example~\cite{luo_light-induced_2021,sie_ultrafast_2019,Sentef15,Hubener17,Schuler20}. 

Furthermore, having access to all three independent dichroic observables -- FDAD, LDAD, and CDAD -- allows to retrieve the phase of the photoemission matrix elements and thus the full complex matrix elements. In this sense, our work can be seen as going towards the first condensed matter “complete” photoionization experiments, in which one obtains the full complex photoemission dipole matrix element, which is already established as the grail of a photoionization experiment in atomic and molecular physics~\cite{Cherepkov_05,hockett_14,Villeneuve_17}.  
In the case of two relevant bands -- as demonstrated for WSe$_2$ close to the maximum of the top valence band -- this phase sensitivity can, in principle, even be exploited to reconstruct the pseudospin texture (or, equivalently, the complex band eigenvectors). We remark that the situation where only two bands in the vicinity of high-symmetry points are relevant for the topological properties (even though they are multi-band systems) covers a large class of materials, including Dirac semimetals, Weyl semimetals~\cite{lv_experimental_2015}, and many two-dimensional topological insulators~\cite{kou_two-dimensional_2017}. In addition, combining our multimodal dichroic approach with photon-energy tunable source (\textit{e.g.} at synchrotron facilities) could allow investigating the full orbital pseudospin and Bloch wavefunction in 3D materials', by extracting the photon-energy dependence of the dichroism ($k_z$ dependence). 

While the orbital pseudospin reconstruction presented here is not perfect and would require more theory input to improve the photoemission matrix elements, it demonstrates that the different dichroic observables are truly independent and serve as a fingerprint of the Bloch wavefunction. A theory that does not include the correct Wannier orbitals or TB Hamiltonian will not yield to the correct phase of the photoemission matrix elements and thus the experimental dichroism in ARPES (LDAD, CDAD, and FDAD), even though the band structure might match. This sensitivity is expected to be pronounced also in multiband systems.
Similarly, as underpinned by the controversy about the Dirac semimetal candidate Cd$_3$As$_2$~\cite{liu_stable_2014,akrap_magneto-optical_2016}, standard ARPES alone can not always distinguish different topological states. The joint experimental and theoretical machinery that we introduced is an important step to bridge this gap. The combination of our novel dichroic observables - FDAD - and theory can be applied to solve open questions about the detailed (topological) nature of electronic structures of some solids, which are still under debate, \textit{e.g.} 1T'-WTe$_2$~\cite{tang_quantum_2017}, topological metals without gaps~\cite{muechler_topological_2016,liu_stable_2014,akrap_magneto-optical_2016}, or light-induced topological phase transitions in graphene and Weyl semimetals~\cite{Oka09, sie_ultrafast_2019}.  

\section*{Acknowledgments}
This work was funded by the Max Planck Society, the European Research Council (ERC) under the European Union’s Horizon 2020 research and innovation program (Grant No. ERC-2015-CoG-682843 and H2020-FETOPEN-2018-2019-2020-01 (OPTOLogic - grant agreement No. 899794)), the German Research Foundation (DFG) within the Emmy Noether program (Grant No. RE 3977/1), the Collaborative Research Center/Transregio 227  "Ultrafast Spin Dynamics" (project B07 and A09) and the Priority Program SPP 2244 (project No.~443366970), and the U.S. Department of Energy (DOE), Office of Basic Energy Sciences, Division of Materials Sciences and Engineering, under contract DE-AC02-76SF00515. T. P. acknowledges financial support from the Alexander von Humboldt Fellowship program of the Alexander von Humboldt Stiftung. M.S. thanks the Alexander von Humboldt Foundation for its support with a Feodor Lynen scholarship and the Swiss National Science Foundation SNF for its support with an Ambizione grant (project No.~193527). S.B. acknowledges financial support from the NSERC-Banting Postdoctoral Fellowships Program. 
 
\section*{Author contributions}
S.B., T.P., and S.D. performed angle-resolved photoemission spectroscopy experiments. S.B. analyzed and interpreted the experimental data. R.E., L.R., and M.W. were responsible for developing the infrastructures allowing these measurements as well as for the overall project direction. M.S. performed the theoretical calculations, their analysis, and interpretation, with the guidance of T.P.D.. M.S. and S.B. wrote the first draft of the manuscript. All authors contributed to the discussions and the final version of the manuscript.

\appendix

\section{Angle-resolved photoemission spectroscopy \label{app:expdetails}}

The angle-resolved photoemission spectroscopy experiments were performed at the Fritz Haber Institute of the Max Planck Society. We used a home-built optical parametric chirped-pulse amplifier (OPCPA) delivering 30 $\mu$J/pulses (800 nm, 30 fs) at 500 kHz repetition rate~\cite{Puppin15}. The second harmonic of the OPCPA output (400 nm) is used to drive high-order harmonic generation (HHG) by tightly focusing (15 $\mu$m FWHM) laser pulses onto a thin and dense Argon gas jet, using a perforated focusing mirror (f=100 mm) with a 1.5 mm hole diameter. The extremely nonlinear interaction between the laser pulses and the Argon atoms leads to the generation of a comb of odd harmonics of the driving laser, extending up to the 11th order. Because the XUV harmonics are generated using an annular driving beam, the copropagating fundamental (400 nm) can be separated from the XUV harmonic beam using a spatial filter (iris) in the far-field. Using this geometry, one can avoid the typically used reflection onto a silicon wafer at Brewster's angle to filter out the energy of the fundamental driving laser, which only works for p-polarized light. Thus, the annular beam HHG scheme allows us to continuously rotate the polarization of the XUV, by simply rotating the polarization of the 400 nm in front of the HHG chamber using a $\lambda/2$-waveplate. Next, a single harmonic (7th order, 21.7 eV) is isolated by reflection off a focusing multilayer XUV mirror and transmission through a 400 nm thick Sn metallic filter. A photon flux of up to 2x10$^{11}$ photons/s at the sample position is obtained (110 meV FWHM)\cite{Puppin19}. The bulk $\mathrm{WSe_2}$ samples are handled by a 6-axis manipulator (SPECS GmbH) and cleaved at a base pressure of 5x10$^{-11}$ mbar. The data are acquired using a time-of-flight momentum microscope (METIS1000, SPECS GmbH), allowing to detect each photoelectron as a single event and as a function of XUV linear polarization angle ($\theta$)~\cite{Medjanik17,maklar20}.

\section{Intensity modulation and phase dependence\label{app:modulation}}

For the experimental geometry displayed on Fig.~\ref{fig:sketch}(a), we can express the $\theta$-dependent polarization vector as $\vec{e}(\theta) = \cos\theta \vec{e}_p + \sin\theta \vec{e}_s$; the unit vector with respect to the $s$ ($p$) polarization, $\vec{e}_s$ ($\vec{e}_p$), are defined by $\vec{e}_s = \vec{e}_y$ and $\vec{e}_p = \cos\beta \vec{e}_x - \sin\beta \vec{e}_z$, where $\vec{e}_r$  ($r=x,y,z$) stands for the corresponding unit vector, and where $\beta=65^\circ$ is the angle of incidence. The photoemission matrix element $M(E,\vec{k},\theta) = \langle \vec{k},E | \vec{e}(\theta) \cdot \hat{\vec{r}} | \psi_{\vec{k}\alpha}\rangle$ is then decomposed into the $s$ and $p$ contributions: $M(E,\vec{k},\theta) = \cos\theta M_p(E,\vec{k}) + \sin\theta M_s(E,\vec{k})$.

From Fermi's Golden rule~\eqref{eq:fermi} we obtain the intensity 
\begin{align}
    \label{eq:intensity_theta}
    I(E,\vec{k},\theta) = |\cos \theta M_p(E,\vec{k}) + \sin \theta M_s(E,\vec{k})|^2 g(E,\vec{k}) \ ,
\end{align}
where $g(E,\vec{k})$ denotes the energy conservation term (in practice, a broadened Dirac delta function). For brevity we will drop the arguments $(E,\vec{k})$ in this appendix. Squaring  the complex matrix element in Eq.~\eqref{eq:intensity_theta} and using trigonometric identities, we obtain
\begin{align}
    \label{eq:intensity_theta_simple}
    I(E,\vec{k},\theta) = g(E,\vec{k})\left(A + B\cos[2\theta - \Phi)] \right) \ ,
\end{align}
where $A = \frac12 (|M_s|^2 + |M_p|^2)$, 
\begin{align*}
    B = \frac12 \left[(|M_p|^2 - |M_s|^2)^2 + 4 \mathrm{Re}[M^*_p M_s] ^2\right]^{1/2}
\end{align*}
and   
\begin{align*}
    \tan \Phi &= 2\frac{\mathrm{Re}\left[M^*_s M_p \right]}{|M_s|^2 - |M_p|^2} \ .
\end{align*}
To see how the phase of the matrix elements enters we introduce $M_{s,p} = |M_{s,p}| e^{i \phi_{s,p}}$, $x = |M_s| / |M_p|$. In terms of these quantities, the phase $\Phi$ is determined by
\begin{align}
    \label{eq:tan2phi}
    \tan \Phi = 2 \cos(\phi_p-\phi_s) \frac{x}{1-x^2}
\end{align}
From Eq.~\eqref{eq:tan2phi} we see that the phase $\Phi$ is determined by (i) the relative strength of the photoemission in $s$ and $p$ direction, respectively, and (ii) the phase difference $\phi_p - \phi_s$. The phase difference is particularly sensitive to the phase of the initial state (including the orbital character and hybridization), thus establishing a connection between the Bloch wavefunction and the phase shift $\Phi$.

\section{Details on the theoretical modeling \label{app:tbdetails}}

\subsection{Tight-binding model\label{app:tbmodel}}
The electronic structure is described by the three-band tight-binding (TB) model from ref.~\cite{liu_three-band_2013} comprising the $|d_{z^2}\rangle$, $|d_{x^2-y^2}\rangle$ and the $|d_{xy}\rangle$ orbitals at the tungsten sites. For convenience we perform the rotation to the magnetic basis, using the $|d_{z^2}\rangle$ (magnetic quantum number $m=0$) and the $|d_{\pm 2}\rangle$ (magnetic quantum number $m=\pm 2$) as basis functions. Using the Wannier representation~\eqref{eq:wannier}, the Bloch wavefunction of the top valence band $\alpha=v$ (omitting the spin state) is then approximated as
\begin{align}
    \label{eq:blochtb}
    \psi_{\vec{k}\alpha}(\vec{r}) &= \frac{1}{N}\sum_{m=0,\pm 2} C_{m}(\vec{k}) \sum_{\vec{R}} e^{i \vec{k}\cdot \vec{R}} w_m(\vec{r} - \vec{R}) \ ,
\end{align}
where the coefficients $C_m(\vec{k})$ are obtained from the corresponding eigenvector of the TB Hamiltonian $H(\vec{k})$. The hybrid orbital for the top valence band is obtained from 
\begin{align}
  \label{eq:hybrid_tb}
  \phi_{\vec{k}}(\vec{r}) = \sum_{m=0,\pm 2} C_{m}(\vec{k}) w_m(\vec{r}) \ .
\end{align}
The Wannier functions $w_m(\vec{r})$ are approximated by the simple atomic orbitals $w_m(\vec{r}) = R_m(r) Y_{2,m}(\hat{\vec{r}})$ ($Y_{l,m}(\hat{\vec{r}})$ denotes the spherical harmonics). We use the same radial dependence as in ref.~\cite{beaulieu_revealing_2020-1}, where the TB model and parameterization of the orbitals has been benchmarked against first-principle calculations. Fig.~\ref{fig:hybrid}(b) shows the Wannier orbitals $w_m(\vec{r})$ used for all calculations, while Fig.~\ref{fig:hybrid}(c) depicts the hybrid orbital constructed from Eq.~\eqref{eq:hybrid_tb}. The bands of the TB model have been shifted to match the ionization potential $\mathrm{I.P.}=4.87$~eV~\cite{rawat_comprehensive_2018}.

\subsection{Photoemission matrix elements\label{app:tbmel}}
In general, the photoemission matrix element with respect to the polarization $\vec{e}$ in the dipole gauge is defined as
\begin{align}
    \label{eq:mel_dip}
    M(E,\vec{k}) = \langle \vec{k},E | \vec{e}\cdot \hat{\vec{r}} | \psi_{\vec{k}\alpha} \rangle = -i \cdot \vec{e} \langle \widetilde{\chi}_{\vec{k},E} | \nabla_{\vec{k}} u_{\vec{k}\alpha} \rangle \ .
\end{align}
Here, $u_{\vec{k}\alpha}(\vec{r}) = e^{-i \vec{k}\cdot\vec{r}} \psi_{\vec{k}\alpha}(\vec{r})$ is the cell-periodic Bloch function, while $\widetilde{\chi}_{\vec{k},E}(\vec{r}) =e^{-i \vec{k}\cdot\vec{r}} \langle \vec{r} | \vec{k},E \rangle $ denotes the cell-periodic photoelectron state. Note that the dipole operator $\hat{\vec{r}}$ is, in principle, ill-defined (unless expressed in terms of Wannier functions~\cite{schuler_gauge_2021}). However, both the initial and the final state are eigenstates of the same Bloch Hamiltonian, which allows to define the dipole matrix element in terms of the Berry connection~\cite{Resta_Quantum-Mechanical_1998,Bianco_Mapping_2011}. We have exploited this relation on the right-hand side of Eq.~\eqref{eq:mel_dip}. Inserting the Wannier representation~\eqref{eq:blochtb} and using $\widetilde{\chi}_{\vec{k},E}(\vec{r} + \vec{R}) = \widetilde{\chi}_{\vec{k},E}(\vec{r})$, the matrix element~\eqref{eq:mel_dip} is found to comprise two contributions $M(E,\vec{k}) = M^\mathrm{dip}(E,\vec{k}) + M^\mathrm{wc}(E,\vec{k})$ with
\begin{subequations}
\begin{equation}
    \label{eqs:mel_dip_dip}
    M^\mathrm{dip}(E,\vec{k}) = \sum_{m} C_m(\vec{k}) \int\!d \vec{r}\, e^{-i \vec{k}\cdot\vec{r}} \widetilde{\chi}^*_{\vec{k},E}(\vec{r}) \vec{e}\cdot \vec{r} w_m(\vec{r}) \ ,
\end{equation}
\begin{equation}
\label{eqs:mel_dip_wc}
    M^\mathrm{wc}(E,\vec{k}) = -i \vec{e}\cdot \sum_{m} \nabla_{\vec{k}}C_m(\vec{k}) \int\!d \vec{r}\, e^{-i \vec{k}\cdot\vec{r}} \widetilde{\chi}^*_{\vec{k},E}(\vec{r}) w_m(\vec{r}) \ .
\end{equation}
\end{subequations}
The first contribution captured by Eq.~\eqref{eqs:mel_dip_dip} captures local dipole transitions, while the second contribution (Eq.~\eqref{eqs:mel_dip_wc}) describes a moving Wannier center as function of $\vec{k}$. The latter term becomes important if the Bloch wave-function has a momentum-dependent weight on separate atoms in the unit cell. Since the wave-function of the top valence band is accurately reproduced by the Wannier representation~\eqref{eq:blochtb} in the subspace of the W $d$ orbitals, $M^\mathrm{wc}(E,\vec{k})$ can be neglected. We performed test calculations to corroborate this argument.

The final states are approximated by plane waves (PW) in what follows, i.\,e. $\widetilde{\chi}_{\vec{k},E}(\vec{r})\approx e^{-i k_\perp z}$. The out-of-plane momentum $p_\perp$ is determined by the kinetic energy of the final state $E = \vec{k}^2/2 + k^2_\perp/2$. In general, the PW approximation is known for being qualitatively accurate for photon energies in the XUV regime~\cite{day_computational_2019,schuler_local_2020-1}, albeit only case-by-case check ensures the predictive power. For WSe$_2$ we have benchmarked the PW approximation against first-principle calculations in ref.~\cite{beaulieu_revealing_2020-1}.

For convenience, we also introduce the atomic photoemission matrix elements
\begin{align}
\label{eq:mel_atomic}
M^{s,p}_m(E,\vec{k}) = \int\!d \vec{r}\, e^{-i \vec{k}\cdot\vec{r}} e^{-i k_\perp z} \vec{e}_{s,p} \cdot \vec{r} w_m(\vec{r}) \ ,
\end{align}
which we evaluate by expanding the plane-wave final state in terms of spherical harmonics. The matrix elements~\eqref{eq:mel_atomic} are defined for $s$ ($p$) polarized light (see Appendix~\ref{app:modulation}). 
From the atomic matrix elements~\eqref{eq:mel_atomic} we can calculate the matrix elements with respect to the initial Bloch state by
\begin{align}
    \label{eq:mel_bloch}
    M^{s,p}(E,\vec{k}) = \sum_{m=0,\pm 2} C_{m}(\vec{k}) M^{s,p}_m(E,\vec{k}) \ .
\end{align}
All quantities discussed in the main text -- circular dichroism and the Fourier signal -- can be expressed in terms of the matrix elements~\eqref{eq:mel_bloch}. 

Representing the dipole operator in terms of the position operator $\hat{\vec{r}}$ (dipole gauge) has several advantages over choosing the momentum operator $\hat{\vec{p}}$ (velocity gauge), which are discussed in the supplemental materials~\cite{supplement}.

\section{Intensity modulation in the atomic limit \label{app:atomiclim}}

It is instructive to compare the modulation of the photoemission intensity upon varying $\theta$ for the underlying orbitals separately. To this end, we computed the orbital-resolved intensity 
\begin{align}
  \label{eq:modulation_orb}
  I_m(E,\vec{k},\theta) = |\cos\theta M^{p}_m(E,\vec{k}) + \sin\theta M^{s}_m(E,\vec{k})|^2 g(E,\vec{k}) \ ,
\end{align}
where we inserted the atomic matrix elements~\eqref{eq:mel_atomic}. Any interference between the orbitals is thus absent in Eq.~\eqref{eq:modulation_orb}. 

\begin{figure}[ht]
\centering
\includegraphics[width=\columnwidth]{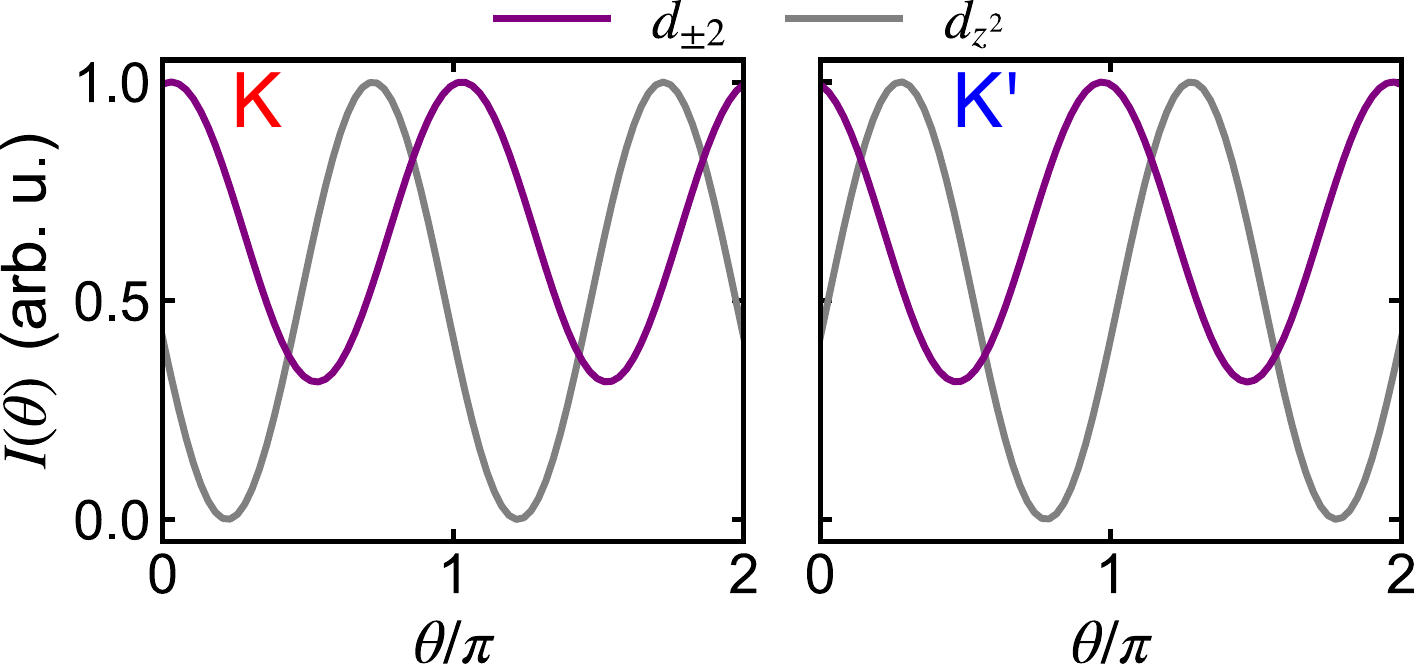}
\caption{Valley-integrated orbital-resolved photoemission intensity $I_m(\theta)$ (analogous to Fig.~\ref{fig:theta_dep}) around the K and K$^\prime$ point, respectivelye (same as in Fig.~\ref{fig:theta_dep}(a)), for $m$ denoting the $d_{\pm 2}$ or the $d_{z^2}$ orbital. The intensity has been normalized to the respective maximum.
\label{fig:theta_atomic}}
\end{figure}

Fig.~\ref{fig:theta_atomic} shows the orbitals-resolved valley-integrated intensity as a function of the polarization angle $\theta$ at $E - E_\mathrm{VBM}=-0.2$~eV (there is only a very weak dependence on $E$ arising from the out-of-plane component $k_\perp$). The dependence for the $d_{\pm 2}$ orbital can be understood intuitively: they are oriented in the $x$--$y$ plane, hence the photoemission probability is maximized if the polarization is pointing out of the plane ($\theta=0$, $p$-polarized light). This behavior is opposite for the $|d_{z^2}\rangle$ orbital, albeit there is a phase shift arising from the geometry. The sign of the modulation associated with the $|d_{z^2}\rangle$ orbital is opposite at the K and K$^\prime$ points, while there is no change for $|d_{\pm 2}\rangle$. Furthermore, the intensity modulation is more pronounced for the $|d_{z^2}\rangle$ orbital. We also note that there is no change upon rotation by $60^\circ$. We conclude that the intensity modulation observed in the experiment (Fig.~\ref{fig:theta_dep}) is due to the interplay of $|d_{z^2}\rangle$ and $|d_{\pm 2}\rangle$ orbitals, and the change of the Bloch state upon rotation manifests as a phase shift.

\section{Reconstruction of the orbital pseudospin \label{app:reconstruct}}

The orbital pseudospin completely determines the photoemission signal, including the circular dichorism, the linear dichroism, and the Fourier signal. Close to the valence band maximum ($\vec{k}\approx \mathrm{K}$ or $\vec{k}\approx \mathrm{K}^\prime$) the hybrid orbital~\eqref{eq:hybrid_tb} is well approximated by $|\phi^{\mathrm{K}, \mathrm{K}^\prime}_{\vec{k}} \rangle = C_0(\vec{k}) | d_{z^2} \rangle + C_{\pm 2}(\vec{k}) |d_{\pm 2}\rangle$. Hence, the Bloch state ~\eqref{eq:blochtb} simplifies to
\begin{align}
  |\psi^{\mathrm{K}, \mathrm{K}^\prime}_{\vec{k}}\rangle \approx \frac{1}{N}\sum_{\vec{R}} e^{i \vec{k}\cdot \vec{R}}\left(C_0(\vec{k}) w_0(\vec{r} - \vec{R}) + C_{\pm 2}(\vec{k}) w_{\pm 2}(\vec{r}-\vec{R}) \right) \ .
\end{align}
The corresponding orbital pseudospin is defined by $\sigma^{\mathrm{K}, \mathrm{K}^\prime}_\nu(\vec{k}) = \langle \psi^{\mathrm{K}, \mathrm{K}^\prime}_{\vec{k}\alpha} | \hat{\sigma}_\nu | \psi^{\mathrm{K}, \mathrm{K}^\prime}_{\vec{k}\alpha} \rangle$. 
The one-to-one correspondence of the complex coefficients $C_m(\vec{k})$ and the orbital pseudospin is given by
\begin{subequations}
\label{eq:sigma_coeff}
\begin{equation}
        \sigma^{\mathrm{K}, \mathrm{K}^\prime}_x(\vec{k}) = 2\mathrm{Re}[C^*_{\pm 2}(\vec{k})C_0(\vec{k})] \ ,
\end{equation}
\begin{equation}
        \sigma^{\mathrm{K}, \mathrm{K}^\prime}_y(\vec{k}) = 2\mathrm{Im}[C^*_{\pm 2}(\vec{k})C_0(\vec{k})] \ ,
\end{equation}
\begin{equation}
        \sigma^{\mathrm{K}, \mathrm{K}^\prime}_z(\vec{k}) = |C_{\pm 2}(\vec{k})|^2 -  |C_0(\vec{k})|^2 \ .
\end{equation}
\end{subequations}
Now we relate the photoemission signal to the pseudospin via Eq.~\eqref{eq:sigma_coeff}. We start from the circular dichroism, which is given by (cf. Eq. (5) in the main text) as
\begin{align}
    \label{eq:cdad}
    I_\mathrm{CD}(E, \vec{k}) = -2 \mathrm{Im}[(M^s(E,\vec{k}))^* M^p(E,\vec{k})] g(E,\vec{k}) \ .
\end{align}
Here $g(E,\vec{k})$ contains the energy conservation. In theory, this factor reduces to a Dirac delta function, but for practical calculations, we replace it by a Gaussian function when calculating the CDAD~\eqref{eq:cdad} (or any other intensity). 

Inserting Eq.~\eqref{eq:mel_bloch} and expressing the complex products of the coefficients in terms of the pseudospin via Eqs.~\eqref{eq:sigma_coeff}, one obtains the linear expression
\begin{align}
    \label{eq:cdad_lin}
    I_\mathrm{CD}(E,\vec{k}) &= \left(\sum_{\nu=1,2,3} A^{\mathrm{K},\mathrm{K}^\prime}_{\mathrm{CD},\nu}(E,\vec{k}) \sigma^{\mathrm{K}, \mathrm{K}^\prime}_\nu(\vec{k}) + B^{\mathrm{K},\mathrm{K}^\prime}_{\mathrm{CD}}(E,\vec{k}) \right) \nonumber \\  &\quad\quad \times g(E,\vec{k}) \ .
\end{align}
The coefficients $A^{\mathrm{K},\mathrm{K}^\prime}_{\mathrm{CD},\nu}$ and the source terms $B^{\mathrm{K},\mathrm{K}^\prime}_{\mathrm{CD}}(E,\vec{k})$ are defined in the supplemental materials~\cite{supplement}.

The expression~\eqref{eq:cdad_lin} is generic -- any intensity can be expressed in a similar linear form with respect to the pseudospin. Following the analogous route for the real part of the Fourier signal, we
find
\begin{align}
    \label{eq:rei2_lin}
    \mathrm{Re}[I_2(E,\vec{k})] & = \left(\sum_{\nu=1,2,3} A^{\mathrm{K},\mathrm{K}^\prime}_{\mathrm{R},\nu}(E,\vec{k}) \sigma^{\mathrm{K}, \mathrm{K}^\prime}_\nu(\vec{k}) + B^{\mathrm{K},\mathrm{K}^\prime}_{\mathrm{R}}(E,\vec{k}) \right)
    \nonumber \\  &\quad\quad
     \times g(E,\vec{k}) \ ,
\end{align}
Finally, we express the imaginary part of the Fourier signal as
\begin{align}
    \label{eq:imi2_lin}
    \mathrm{Im}[I_2(E,\vec{k})] &= \left(\sum_{\nu=1,2,3} A^{\mathrm{K},\mathrm{K}^\prime}_{\mathrm{I},\nu}(E,\vec{k}) \sigma^{\mathrm{K}, \mathrm{K}^\prime}_\nu(\vec{k}) + B^{\mathrm{K},\mathrm{K}^\prime}_{\mathrm{I}}(E,\vec{k}) \right)
    \nonumber \\  &\quad\quad
     \times g(E,\vec{k}) \ .
\end{align}
The terms in Eq.~\eqref{eq:rei2_lin} and \eqref{eq:imi2_lin} are defined in the supplemental materials~\cite{supplement}.

Summarizing Eqs.~\eqref{eq:cdad_lin}--\eqref{eq:imi2_lin}, we can express the three quantities $I_\mathrm{CD}(E,\vec{k})$, $\mathrm{Re}[I_2(E,\vec{k})]$, and $\mathrm{Im}[I_2(E,\vec{k})]$ as linear function of the pseudospin, which can conveniently be cast into the system of equations
\begin{widetext}
\begin{align}
    \label{eq:mastereq}
    \begin{bmatrix} I_\mathrm{CD}(E,\vec{k}) \\ \mathrm{Re}[I_2(E,\vec{k})] \\ \mathrm{Im}[I_2(E,\vec{k})]  \end{bmatrix}
    = \begin{bmatrix} \mathcal{A}^{\mathrm{K},\mathrm{K}^\prime}_{\mathrm{CD},x}(E,\vec{k}) & \mathcal{A}^{\mathrm{K},\mathrm{K}^\prime}_{\mathrm{CD},y}(E,\vec{k}) & \mathcal{A}^{\mathrm{K},\mathrm{K}^\prime}_{\mathrm{CD},z}(E,\vec{k}) \\
    \mathcal{A}^{\mathrm{K},\mathrm{K}^\prime}_{\mathrm{R},x}(E,\vec{k}) & \mathcal{A}^{\mathrm{K},\mathrm{K}^\prime}_{\mathrm{R},y}(E,\vec{k}) & \mathcal{A}^{\mathrm{K},\mathrm{K}^\prime}_{\mathrm{R},z}(E,\vec{k}) \\
    \mathcal{A}^{\mathrm{K},\mathrm{K}^\prime}_{\mathrm{I},x}(E,\vec{k}) & \mathcal{A}^{\mathrm{K},\mathrm{K}^\prime}_{\mathrm{I},y}(E,\vec{k}) & \mathcal{A}^{\mathrm{K},\mathrm{K}^\prime}_{\mathrm{I},z}(E,\vec{k})
    \end{bmatrix}
     \begin{bmatrix}
     \sigma^{\mathrm{K}, \mathrm{K}^\prime}_x(\vec{k}) \\  \sigma^{\mathrm{K}, \mathrm{K}^\prime}_y(\vec{k}) \\  \sigma^{\mathrm{K}, \mathrm{K}^\prime}_z(\vec{k})
    \end{bmatrix}
    + \begin{bmatrix}
     \mathcal{B}^{\mathrm{K},\mathrm{K}^\prime}_{\mathrm{CD}}(E,\vec{k}) \\ \mathcal{B}^{\mathrm{K},\mathrm{K}^\prime}_{\mathrm{R}}(E,\vec{k}) \\ \mathcal{B}^{\mathrm{K},\mathrm{K}^\prime}_{\mathrm{I}}(E,\vec{k})
    \end{bmatrix} \ ,
\end{align}
where we have abbreviated $\mathcal{A}^{\mathrm{K},\mathrm{K}^\prime}_{r,\nu}(E,\vec{k}) = g(E,\vec{k}) A^{\mathrm{K},\mathrm{K}^\prime}_{r,\nu}(E,\vec{k})$ and $\mathcal{B}^{\mathrm{K},\mathrm{K}^\prime}_{r}(E,\vec{k}) = g(E,\vec{k}) B^{\mathrm{K},\mathrm{K}^\prime}_{r}(E,\vec{k})$ ($r=\mathrm{CD},\mathrm{R},\mathrm{I}$).

\end{widetext}

To construct the coefficient matrix $\mathcal{A}^{\mathrm{K},\mathrm{K}^\prime}_{r,\nu}(E,\vec{k})$ and the source term $\mathcal{B}^{\mathrm{K},\mathrm{K}^\prime}_{r}(E,\vec{k})$ only two ingredients are required: (i) the atomic matrix elements~\eqref{eq:mel_atomic}, and (ii) the energy conservation $g(E,\vec{k})$. The atomic matrix elements are mostly determined by the orbital symmetry (angular momentum), which can be guessed from the crystal structure or obtained from first principles. 
The energy conservation factor $g(E,\vec{k})$ can be extracted from experimental spectra by fitting a Gaussian function as function of $E$ at every momentum point $\vec{k}$.

With all terms (except of the pseudospin vector) on the right-hand side of Eq.~\eqref{eq:mastereq} determined, Eq.~\eqref{eq:mastereq} can be solved for $\sigma^{\mathrm{K},\mathrm{K}^\prime}_\nu(\vec{k})$. For a fixed energy $E$ this is possible for all $\vec{k}$ with sufficient signal. The determinant of the coefficient matrix $\mathcal{A}^{\mathrm{K},\mathrm{K}^\prime}_{r,\nu}(E,\vec{k})$ is proportional to $g(E,\vec{k})$; thus we solve Eq.~\eqref{eq:mastereq} only momenta obeying $g(E,\vec{k}) > \epsilon$. Normalizing $g(E,\vec{k})$ to one, we fix $\epsilon=10^{-1}$.

We have tested the self-consistency within the theory by calculating the left-hand side of Eq.~\eqref{eq:mastereq} and solving for the pseudospin. Comparing the thus obtained solution to the directly calculated pseudospin (via Eq.~\eqref{eq:sigma_coeff}) yields perfect agreement. 
We repeated the procedure adding small random noise to the input signal; the reconstructed pseudospin is still in excellent agreement with the calculated texture. 

For reconstructing the pseudospin from experimental data -- as presented in the main text -- a direct solution of Eq.~\eqref{eq:mastereq} in terms of matrix inversion gives rise to artifacts; most importantly, the normalization of the pseudospin
\begin{align}
    \label{eq:pseudo_norm}
    \sigma^{\mathrm{K},\mathrm{K}^\prime}_x (\vec{k})^2 +  \sigma^{\mathrm{K},\mathrm{K}^\prime}_y (\vec{k})^2 +  \sigma^{\mathrm{K},\mathrm{K}^\prime}_z (\vec{k})^2 = 1
\end{align}
is violated. Therefore, we switch to the more least-square fitting algorithm. Furthermore, we constrain the solution by the normalization condition~\eqref{eq:pseudo_norm} by adding a penalty term, which is chosen to ensure Eq.~\eqref{eq:pseudo_norm} is obeyed up to $10^{-3}$. Following this procedure yields the pseudospin textures presented in the main text.


%

\end{document}